\documentclass[10pt,aps,prb,showpacs,superscriptaddress,twocolumn,
floatfix]{revtex4}
\usepackage{amsmath}
\usepackage{citesort}
\usepackage{graphicx}
\begin{document}
\title{Optical spectral weight distribution in $d$-wave
  superconductors}
\author{J.P. Carbotte}
\affiliation{Department of Physics and Astronomy, McMaster University,\\
Hamilton, Ontario, L8S 4M1 Canada}
\author{E. Schachinger}
\email{schachinger@itp.tu-graz.ac.at}
\homepage{www.itp.tu-graz.ac.at/~ewald}
\affiliation{Institute for Theoretical and Computational Physics,\\
    Graz University of Technology, A-8010 Graz, Austria}
\date{\today}
\begin{abstract}
  The distribution in frequency of optical spectral weight
  remaining under the real part of the optical conductivity
  in the superconducting state of a $d$-wave superconductor
  depends on impurity concentration,
  on the strength of the
  impurity potential as well as on temperature and
  there is some residual absorption even at $T=0$. In BCS theory
  the important weight is confined to the microwave region
  if the scattering is sufficiently weak.
  In an Eliashberg formulation substantial additional weight
  is to be found in the incoherent, boson assisted background
  which falls in the infrared and is not significantly depleted by the
  formation of the condensate, although it is shifted as a result
  of the opening of a superconducting gap.
\end{abstract}
\maketitle
\newpage
\section{Introduction}

When a metal enters its superconducting state, optical spectral
weight is lost at finite frequencies
under the real part of the optical conductivity,
$\sigma_1(T,\omega)$.\cite{mars1}
Provided the change in kinetic energy between normal and
superconducting state is small and can be neglected, the
missing spectral weight reappears as a contribution at zero
frequency
which originates in the superfluid, and the
over all optical sum rule of Ferrell, Glover, and Tinkham\cite{%
ferrell,tinkham} remains unchanged. The distribution
in frequency of the remaining spectral weight under
$\sigma_1(\omega)$ $(\omega>0)$ depends on gap symmetry,
on the nature of the inelastic scattering involved,
on the concentration and 
scattering strength of the impurities,
and on temperature.\cite{vM}
In this paper we consider explicitely the
case of $d$-wave gap symmetry
within a generalized Eliashberg formalism.\cite{schach8}
In this approach the optical conductivity (as well as the
quasiparticle spectral density) contains
an incoherent part associated with boson assisted absorption
which is not centered about zero frequency and
which contributes to the optical spectral weight in the
infrared range. In addition there is the usual quasiparticle
contribution of BCS theory.
Alternate approaches to include inelastic scattering exist. In
several works, the quasiparticle scattering rate due to coupling
to spin fluctuations is simply
added to a BCS formalism through an additional scattering channel.%
\cite{quinla,hirschf,quinlb,hirschf1} Nevertheless,
whenever we refer to BCS within this paper we mean
the standard theory without these additional features.

In BCS theory
the London penetration depth\cite{BCS6,carb} at zero
temperature $[\lambda_L(0)]$ in the clean limit is given by
$\lambda_L^{-2}(0) = \lambda^{-2}_{cl}(0) = 4\pi ne^2/m =%
 \Omega^2_p$ ($n$ is the
free electron density, $e$ is the charge on the electron,
$m$ is its mass, $\Omega_p$ is the plasma frequency,
and we have set the velocity of light equal to 1)
and all the optical spectral
weight condenses. However, as the impurity mean free path is reduced,
not all the spectral
weight is transferred to the condensate\cite{tanner,liu} and there remains
some residual impurity induced absorption.\cite{turner,corson,resabs}
Details depend on gap symmetry.

In Eliashberg theory the pairing interaction is
described by an electron-phonon spectral density, denoted by
$\alpha^2F(\omega)$.\cite{BCS6,carb,no7}
Twice the first inverse
moment of $\alpha^2F(\omega)$,
gives the quasiparticle mass renormalization
with the effective $(m^\ast)$ to bare $(m)$ mass
ratio $m^\ast/m = 1+\lambda$.
While the gap and renormalization function of Eliashberg theory
acquire a frequency dependence which requires numerical
treatment, a useful, although not exact, approximation is
to assume that the important frequencies in $\alpha^2F(\omega)$
are much higher than the superconducting energy scale and, thus,
one can
approximate the renormalizations by a constant $\lambda$ value.%
\cite{carb}
In this approximation, the zero temperature
penetration depth is $\lambda^{-2}_L(0)\simeq(4\pi ne^2/mc^2)%
[1/(1+\lambda)]$ in the clean limit.
Thus, the electron-phonon renormalization
simply changes the bare mass in the London expression to the
renormalized mass $m^\ast$. This result 
does not depend explicitly on the gap and 
holds independent of its symmetry.
A naive interpretation of
this result is that only the
coherent quasiparticle part of the electron-spectral density
[which contains approximately $1/(1+\lambda)$ of the total
spectral weight of one] condenses.
While this is approximately true, we will see that the
incoherent part which contains the remaining $\lambda/(1+\lambda)$
part of the spectral weight
is also involved, although in a more minor and subtle way.

In an $s$-wave superconductor the entire incoherent
part of the conductivity is shifted upward by twice the gap value,
$\Delta$, when compared to its normal state. It is also slightly
distorted but, to a good approximation, it remains
unchanged. The fact that there is a $2\Delta$ shift
between normal and superconducting state
implies that an optical spectral weight shift originates from
this contribution even if its overall contribution
to the sum rule should remain the same.
For a $d$-wave superconductor the situation
is more complex because the gap is anisotropic and, thus, the
shift by $2\Delta(\phi)$ varies with the polar angle
$\phi$  on the two-dimensional Fermi surface of the CuO$_2$ planes.

The goal of this paper is to understand, within an Eliashberg
formalism, how the remaining area under the real part of the
optical conductivity is distributed in frequency, how this
distribution is changed by finite temperature effects and
by the introduction of elastic impurity scattering, and
what information can be obtained from such studies about the
superconducting state and the nature of the mechanism which
drives it.

In reference to $d$-wave superconductivity in the cuprates two boson
exchange models which have received much attention
are the Nearly Antiferromagnetic Fermi Liquid (NAFFL) model%
\cite{pines1,pines2,schach4,schach5,schach7,schach6} and
the Marginal Fermi Liquid (MFL) model.\cite{varma1,varma2,varma3}
Both models
are characterized by an appropriate charge carrier-exchange boson spectral
density $I^2\chi(\omega)$ which replaces the $\alpha^2F(\omega)$
of the phonon case\cite{BCS6,mcmillan,mcmillan1,mars4}
and which reflects the nature of the inelastic scattering envisioned.
In the NAFFL model a
further complication arises in that we would expect
$I^2\chi(\omega)$ to be very anisotropic as a function
of momentum on the Fermi surface. For simplicity we ignore this
complication here. Also, in principle, a
different spectral weight function can enter the gap and renormalization
channel, respectively.

In Section~\ref{sec:2}, we provide some theoretical background. The
quasiparticle spectral density as a function of energy is
considered as is the effect of impurities on it.
In Sec.~\ref{sec:3}  we give the necessary formulas for the optical
conductivity and discuss some results. In Sec.~\ref{sec:4} the
conditions under which a partial sum rule involving only the
quasiparticle part of the spectral density
can be expected are described.
Section~\ref{sec:5} deals with issues associated with the residual
absorption and Sec.~\ref{sec:6} deals with a more detailed discussion
of optical spectral weight readjustment due to superconductivity.
Conclusions are found in Sec.~\ref{sec:7}.

\section{Quasiparticle spectral density}
\label{sec:2}

We begin with a discussion of the quasiparticle spectral density
which will allow us to understand the basic features expected
of the optical conductivity.
In Nambu notation the $2\times 2$-matrix Green's function
$\hat{G}({\bf k},\omega)$ in the superconducting state is given in
terms of the single quasiparticle dispersion $\varepsilon_{\bf k}$
with momentum {\bf k}, the renormalized Matsubara frequency
$\tilde{\omega}(\omega)$ and the pairing energy
$\tilde{\Delta}_{\bf k}(\omega)$
which for a $d$-wave superconductor is proportional to
$\cos(2\phi)$. In terms of Pauli's $\hat{\tau}$ matrices
\begin{equation}
  \label{eq:gf}
  \hat{G}({\bf k},\omega) = \frac{\tilde{\omega}(\omega)\hat{\tau}_0+
  \varepsilon_{\bf k}\hat{\tau}_3+\tilde{\Delta}_{\bf k}(\omega)
  \hat{\tau}_1}
  {\tilde{\omega}^2(\omega)-\varepsilon_{\bf k}^2-
   \tilde{\Delta}_{\bf k}^2(\omega)}.
\end{equation}
The quasiparticle spectral density $A({\bf k},\omega)$ is given by
\begin{eqnarray}
  A({\bf k},\omega) &=& -\frac{1}{\pi}\Im{\rm m}
    G_{11}({\bf k},\omega+i0^+)\nonumber\\
  &=& -\frac{1}{\pi}\Im{\rm m}\frac{\tilde{\omega}(\omega+i0^+)+
   \varepsilon_{\bf k}}{\tilde{\omega}^2(\omega+i0^+)-
   \varepsilon_{\bf k}^2-\tilde{\Delta}_{\bf k}^2(\omega+i0^+)}.
  \label{eq:sd}  
\end{eqnarray}
The generalized Eliashberg equations applicable to $d$-wave
gap symmetry which include renormalization effects in the
$\omega$-channel have been written down before and
will not be repeated here.\cite{schach8}
They are a set of
coupled non-linear integral equations for $\tilde{\omega}(\omega)$
and $\tilde{\Delta}_{\bf k}(\omega)$ which depend on an
electron-boson spectral density $I^2\chi(\omega)$.
The boson exchange mechanism involved in superconductivity
is what determines its shape in frequency and its magnitude.
In general, the projection of the electron-boson interaction on
the $\tilde{\Delta}$ and $\tilde{\omega}$-channel
can be different but for simplicity, here, the
same form of $I^2\chi(\omega)$ is used in both channels but with
a different magnitude:
we use $gI^2\chi(\omega)$ with $g\ne 1$ for the
$\tilde{\Delta}$-channel.

\begin{figure}[t]
  \includegraphics[width=8cm]{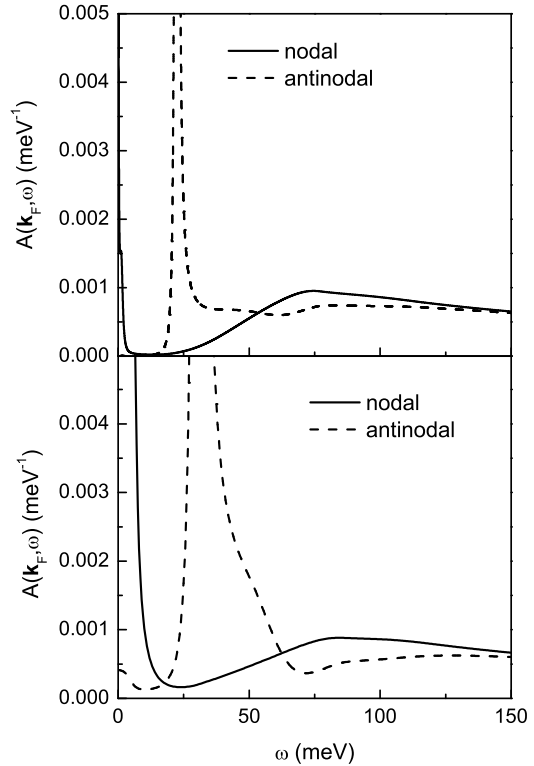}  
  \caption{The charge carrier spectral density $A({\bf k}_F,\omega)$
as a function of $\omega$ for a $d$-wave superconductor based on
the electron-spin fluctuation spectral density $I^2\chi(\omega)$
shown in the inset of \protect{Fig.~\ref{fig:s1}}.
The solid curve
applies to the nodal while the dashed curve is for
the antinodal direction. The top frame is for a pure sample with
impurity parameters $\Gamma^+ = 0.003\,$meV and $c=0.2$ while
the bottom frame is for $\Gamma^+=0.63\,$meV and $c=0$.
}
  \label{fig:spec}
\end{figure}

In Fig.~\ref{fig:spec}  we present numerical results for
$A({\bf k}_F,\omega)$ based on numerical
solutions of the Eliashberg equations.
The kernel $I^2\chi(\omega)$ used for the numerical
work is shown in the inset in the top frame of Fig.~\ref{fig:s1}
and was obtained
from consideration of the infrared optical conductivity of
YBa$_2$Cu$_3$O$_{6.95}$ (YBCO$_{6.95}$).\cite{schach7}
Besides coupling to an optical resonance at $41\,$meV
(the energy where a spin resonance is also seen
in the inelastic neutron scattering\cite{bourges})
which grows with decreasing
temperature into the superconducting state,
there is also additional coupling
to a broad spin fluctuation spectrum background
of the form introduced by Millis {\it et al.}\cite{pines1} in their
NAFFL model. This is seen as the long tail in $I^2\chi(\omega)$
which extends to very high energies of order $400\,$meV.
The existence of these tails is a universal property of the
cuprates.\cite{tanner,liu,schach6,puchkov,homes,tu}
This energy scale is of the order of the magnetic parameter $J$ in
the $t-J$ model.\cite{sorella}
A flat background spectrum is also characteristic of the
MFL model.\cite{varma1,varma2,varma3}
In this work, the shape and size of $I^2\chi(\omega)$
is fixed from our previous fit to optical data\cite{schach7}
and left unchanged.
It applies at low temperatures in the superconducting state
($T\sim 10\,$K).

The top frame of Fig.~\ref{fig:spec} gives results for the
charge carrier spectral density $A({\bf k}_F,\omega)$ vs $\omega$
where ${\bf k}_F$ implies that we consider only the Fermi energy
in Eq.~\eqref{eq:sd}. The results are for a pure sample with
$\Gamma^+=0.003\,$meV and $c=0.2$. Here, $\Gamma^+$
is proportional to the impurity concentration and is related to
the normal state impurity scattering rate $(\tau^{-1}_{imp})$
equal to $2\pi\Gamma^+[1/(c^2+1)]$, where $c = 1/[2\pi N(0) V_{\rm imp}]$.
$N(0)$ is the normal state density of states at the Fermi energy and
$V_{\rm imp}$ the strength of the impurity potential.
These
impurity parameters were determined to fit well the microwave
data in YBCO$_{6.99}$ obtained by Hosseini {\it et al.}\cite{hoss}
The solid curve is for the nodal
direction and the dashed curve for the antinodal direction.
The spectral gap is the value of
$\Delta(\omega+i0^+) = \tilde{\Delta}(\omega+i0^+)/\tilde{\omega}(\omega+i0^+)$
evaluated at the frequency of the coherence peak in the density of states
\begin{eqnarray}
  \frac{N(\omega)}{N(0)} &=& \Re{\rm e}\left\langle\frac{\tilde{\omega}
  (\omega+i0^+)}{\sqrt{\tilde{\omega}^2(\omega+i0^+)-
  \tilde{\Delta}^2(\omega+i0^+)}}\right\rangle'\nonumber\\
  &\equiv&
   \Re{\rm e}\left[\Omega(\omega)\right],
  \label{eq:dos}
\end{eqnarray}
and is equal to $22.3\,$meV. This is also the position of the
large peak seen in the dashed curve in the top frame of Fig.~\ref{fig:spec}.
However, there is no gap in the nodal direction, and in this case
the spectral function is peaked about $\omega=0$. It rapidly
decays to nearly zero within a very narrow frequency range
determined by a combination of the small impurity scattering rate
which we have included and the equally
small inelastic scattering which reflects the presence of
$I^2\chi(\omega)$ and finite temperature. A second peak is also
observed at higher energies but
with reduced amplitude. This peak has its origin
in the incoherent boson assisted processes described by the
spectral density $I^2\chi(\omega)$.
Note that the two contributions are well separated.
In the constant $\lambda$ model, the coherent part
\begin{equation}
  \label{eq:nssd}
  A({\bf k}_F,\omega) = \frac{1}{1+\lambda}\,
  \frac{\pi\Gamma^+/[(1+\lambda)(1+c^2)]}
  {\omega^2+\{\pi\Gamma^+/[(1+\lambda)(1+c^2)]\}^2},
\end{equation}
is a Lorentzian of width $\pi\Gamma^+/[(1+\lambda)(1+c^2)]$ and 
has total weight of $1/(1+\lambda)$. The
remaining weight in the complete spectral density which is normalized
to one, is thus to be found in the incoherent, boson assisted
background. Returning to
the antinodal direction, we see that in this case
the separation between quasiparticle peak and
incoherent boson assisted background is lost as the
two contributions overlap significantly. In the bottom frame of
Fig.~\ref{fig:spec} we show similar results for the charge
carrier spectral density but now a larger amount of
impurity scattering is included with $\Gamma^+=0.63\,$meV
(Ref.~\onlinecite{schach2})
and the unitary limit is taken, i.e. $c=0$. In this instance,
even for the nodal direction, impurities have the effect of
filling in the
region between quasiparticle and incoherent background (solid curve).
Also for the antinodal direction (dashed curve),
because we are in $d$-wave, the region below
the gap energy which is now $\sim 30\,$meV is filled in significantly. It
would be zero in BCS $s$-wave. At $\omega=0$, $\tilde{\omega}(0) =
i\gamma$ and in antinodal direction
\begin{equation}
  \label{eq:ano}
  A({\bf k}_F,\omega=0) = \frac{1}{\pi(1+\lambda)}\,
  \frac{\gamma/(1+\lambda)}{\Delta^2+[\gamma/(1+\lambda)]^2},
\end{equation}
which is finite. Here $\gamma$ is
the quasiparticle scattering rate at zero frequency in the
superconducting state. It is calculated in Sec.~\ref{sec:5}.
This limit is not universal
in contrast to the universal limit found 
by Lee\cite{lee} for the real part of the
electrical conductivity at zero temperature which
is $(ne^2/m)\{1/[\pi\Delta(1+\lambda)]\}$
in the constant $\lambda$ model. Note that what enters
the universal limit is
the renormalized mass $m(1+\lambda) = m^\ast$ rather than the
bare mass. This important fact has generally been overlooked
in the discussion of this limit even though the difference
can be numerically large (order $\sim 3$). We note one technical
point about our
Eliashberg numerical solutions. In all cases $I^2\chi(\omega)$
is kept fixed as is $T_c = 92\,$K. In a $d$-wave superconductor
the introduction of impurities, of course, reduces the critical
temperature. What is done is that the parameter $g$
which multiplies $I^2\chi(\omega)$ in the gap channel is readjusted
slightly to keep $T_c$ fixed. This procedure leads to the larger
value of the spectral gap seen in the bottom frame of
Fig.~\ref{fig:spec} as compared with the top frame (dashed lines).

\section{Infrared conductivity}
\label{sec:3}

A general expression for the infrared optical conductivity
at temperature $T$ in a BCS $d$-wave superconductor is%
\cite{schach1,schach2,schur}
\begin{widetext}
\begin{equation}
  \label{eq:1}
  \sigma(T,\nu) = -\frac{\Omega^2_p}{4\pi}\left\langle\left[
      -\int\limits_0^\infty\!d\omega\,\tanh\left(\frac{\beta\omega}{2}
        \right)J(\omega,\nu)+\int\limits_{-\nu}^\infty\!d\omega\,
        \tanh\left(\beta\frac{\omega+\nu}{2}\right)
        J(-\omega-\nu,\nu)
      \right]\right\rangle,
\end{equation}
where the function $J(\omega,\nu)$ takes on the form
\begin{eqnarray}
  2J(\omega,\nu) &=& \frac{1}{E_1(\omega)+E_2(\omega,\nu)}
    \left[1-N(\omega)N(\omega+\nu)
    -P(\omega)P(\omega+\nu)\right]\nonumber\\
  &&+\frac{1}{E^\ast_1(\omega)-E_2(\omega,\nu)}
    \left[1+N^\ast(\omega)N(\omega+\nu)
    +P^\ast(\omega)P(\omega+\nu)\right].
  \label{eq:2}
\end{eqnarray}
In Eq.~(\ref{eq:1}) $\beta = 1/k_B T$, with $k_B$ the Boltzmann factor.
In Eq.~(\ref{eq:2})
\begin{subequations}
\begin{equation}
  E_1(\omega) = \sqrt{\tilde{\omega}^2(\omega+i0^+)-
  \tilde{\Delta}^2(\omega+i0^+)},
  \qquad E_2(\omega,\nu) = E_1(\omega+\nu),
\end{equation}
and
\begin{equation}
  N(\omega) = \frac{\tilde{\omega}(\omega+i0^+)}{E_1(\omega)},
  \qquad P(\omega) = \frac{\tilde{\Delta}(\omega+i0^+)}{E_1(\omega)},
\end{equation}
\end{subequations}
\end{widetext}
and $E^\ast_1(\omega)$, $N^\ast(\omega)$, and $P^\ast(\omega)$
are the complex conjugates of $E_1(\omega)$, $N(\omega)$, and
$P(\omega)$, respectively.
These expressions hold for an Eliashberg superconductor as well
as for BCS in which case 
the gap $\tilde{\Delta}(\omega)$ does not 
depend on frequency; it only depends on temperature, and on angle.
Here, for brevity we have suppressed these dependencies but they are
implicitly implied by the brackets $\langle\cdots\rangle$ in Eq.~%
(\ref{eq:1}) which denote an angular average over momentum
directions of electrons on the Fermi surface at a given temperature.

\begin{figure}[t]
  \includegraphics[width=9cm]{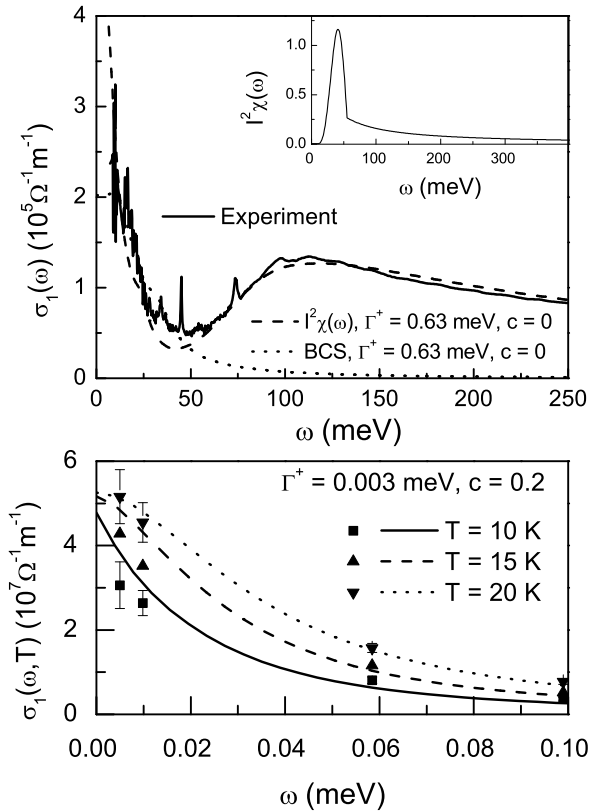}
  \caption{Top frame: Real part of the optical conductivity
   $\sigma_1(T,\omega)$ vs $\omega$ for an optimally doped,
   untwinned YBCO$_{6.95}$ single crystal at $T=10\,$K. The solid
   line represents the experimental data reported by
   Homes {\it et al.}\protect{\cite{homes}}, the dashed line
   the result of a fit to a full Eliashberg calculation using the
   electron-boson spectral density $I^2\chi(\omega)$ shown
   in the inset and the impurity parameters $\Gamma^+ = 0.63\,$meV
   and $c=0$.\protect{\cite{schach2}} The dotted line presents,
   for comparison, the result of a BCS calculation using the same
   impurity parameters.
    Bottom frame: the microwave
   region of $\sigma_1(T,\omega)$ for $\Gamma^+=0.003\,$meV and
   $c=0.2$ which fits well the data of Hosseini {\it et al.}\protect{%
\cite{hoss}} (shown as symbols)
   for three temperatures, $T=10\,$K (solid line), $T=15\,$K
   (dashed line), and $T=20\,$K (dotted line).\protect{\cite{schach1}}
    Again, the
   $I^2\chi(\omega)$ shown in the inset of the top frame has
   been used.
    }
  \label{fig:s1}
\end{figure}

Figure~\ref{fig:s1} presents two fits of theoretical results
to experimental data for the real part of the optical conductivity
$\sigma_1(T,\omega)$. The top frame presents a comparison with
data reported by Homes {\it et al.}\cite{homes} for an untwinned,
optimally doped YBCO$_{6.95}$ single crystal (solid line) at
$T=10\,$K. The dashed line corresponds to the
best fit theoretical results generated using extended Eliashberg
theory. The phenomenologically determined electron-boson spectrum
$I^2\chi(\omega)$ reported by Schachinger {\it et al.}\cite{schach7}
(shown in the inset) was used. The impurity parameters
$\Gamma^+=0.63\,$meV and $c=0$ resulted in this best fit.\cite{schach2}
For comparison the dotted line
corresponds to the results of a BCS calculation using the same
impurity parameters. It is obvious that the BCS calculation
cannot reproduce the boson assisted higher energy incoherent
background which starts at about $80\,$meV. The full Eliashberg
theory, on the other hand, is capable of modeling very well
the experimental $\sigma_1(T,\omega)$ data over the whole infrared region.
The bottom frame of Fig.~\ref{fig:s1} 
shows $\sigma_1(T,\omega)$ restricted to the microwave region
up to $\omega=0.1\,$meV. Three temperatures
are considered, namely $T=10\,$K
(solid curve), $T=15\,$K (dashed curve), and $T=20\,$K (dotted
curve). The impurity parameters were varied to get a
good fit to the data of a high purity YBCO$_{6.99}$ sample
reported by Hosseini {\it et al.}\cite{hoss} and presented
by symbols. The best fit was found
for $\Gamma^+=0.003\,$meV and $c=0.2$. It is clear
that this sample is very pure and that it is not in the unitary limit.
All curves for $\sigma_1(T,\omega)$ vs $\omega$ in this frame
show the upward curvature characteristic
of finite $c$ values. Unitary scattering would give a
downward curvature in disagreement with the data.

The excellent agreement between theory and experiment
shown in Fig.~\ref{fig:s1} encourages us to
apply theory to discuss in detail, issues connected
with the redistribution of optical spectral weight in going
from the normal (not always available in experiment)
to the superconducting state and the effect of temperature and
impurities on it.

\begin{figure}[t]
  \includegraphics[width=9cm]{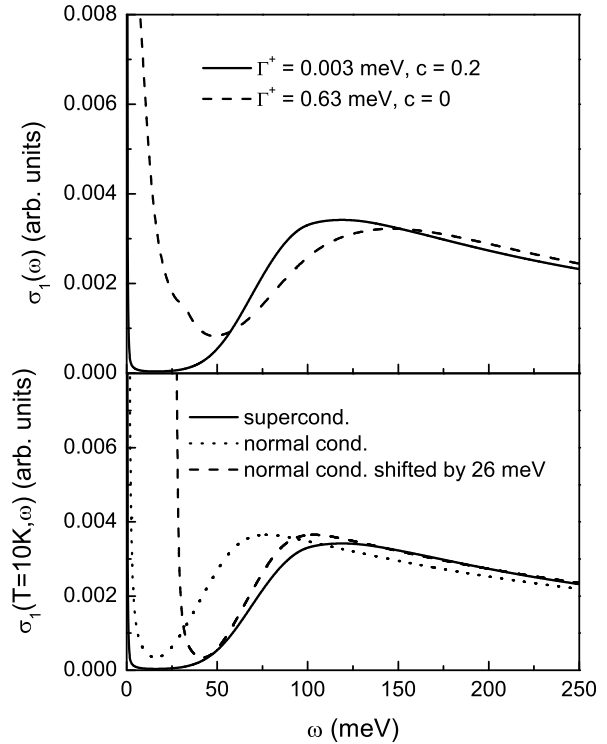}
  \caption{Top frame: Real part of the optical conductivity
   $\sigma_1(\omega)$
   vs $\omega$ in units of $\Omega_p^2/(8\pi)$.
   The solid curve is for a pure sample with
   impurity parameters $\Gamma^+ = 0.003\,$meV, $c=0.2$ and the
   dashed curve is for $\Gamma^+=0.63\,$meV and $c=0$.
   The temperature $T=10\,$K. The
   electron-boson spectral density $I^2\chi(\omega)$ used is
   shown in the inset of the top frame of Fig.~\protect{\ref{fig:s1}}.
   For the solid curve, the narrow coherent quasiparticle
   peak centered at $\omega=0$ is well separated from the higher energy
   incoherent, boson-assisted region. This separation is
   less clear in the dashed curve.
  Bottom frame: Real part of the optical conductivity $\sigma_1(T,\omega)$
  vs $\omega$ in units of $\Omega^2_p/(8\pi)$ for a pure sample with
  impurity parameters $\Gamma^+=0.003\,$meV and $c=0.2$ at
  $10\,$K. The superconducting state (solid line) is compared
  with the normal state, i.e. setting the gap
  $\tilde{\Delta}(\omega)=0$ in the Eliashberg equations (dotted line).
  The dashed curve is
  a repeat of the normal state curve but has been shifted in
  frequency by $26\,$meV.}
  \label{fig:2b}
\end{figure}

The real part of the optical conductivity $\sigma_1(T,\omega)$ as
a function of $\omega$ is shown in the top frame of
Fig.~\ref{fig:2b}. A factor $\Omega_p^2/(8\pi)$
has been omitted from all theoretical
calculations and so $\sigma_1(T,\omega)$ is in meV$^{-1}$. 
In these units the
usual FGT sum rule which gives the total available optical
spectral weight $\int_0^\infty d\omega\,\sigma_1(\omega) = \pi$
(including the superfluid contribution at $\omega=0$).
Two cases are shown in
the frequency range $0^+\le\omega\le 250\,$meV. One is for the very
pure sample (solid curve) with
$\Gamma^+=0.003\,$meV and $c=0.2$. The other is for a less pure sample
(dotted curve) with $\Gamma^+=0.63\,$meV
in the unitary limit, $c=0$.
In the solid curve we clearly see a separate
quasiparticle contribution peaked about $\omega=0$ which
is responsible for a
coherent Drude like contribution to the real part
of the optical conductivity.
In this process the energy of the
photon is transferred to the electrons with the impurities
providing a momentum sink. The width of the quasiparticle peak
and corresponding Drude peak is related
to the impurity scattering rate.
Because we are using Eliashberg theory
there is also a small contribution to this width coming
from the thermal population of
excited spin fluctuations.
In addition, there is a separate incoherent contribution at
higher frequencies.
This second contribution involves the creation of spin fluctuations
during the absorption process. Its shape reflects details of the
frequency dependence of the spectral density $I^2\chi(\omega)$
involved. For the normal state at temperature $T > T_c$ the spectral
density $I^2\chi(\omega)$ in the NAFFL model does not show the
resonance peak seen in the insert of the top frame of Fig.~\ref{fig:s1}
but consists mainly of the reasonably flat background. This implies
that in this region MFL behavior results with
optical and quasiparticle lifetimes linear in frequency
and in temperature. The energy scale associated with this behavior
is the spin fluctuation scale $\omega_{SF}$. This is verified
in numerous experiments in the cuprates
as reviewed by Puchkov {\it et al.}\cite{puchkov}
Just as for the charge carrier spectral
density discussed in the previous section, the optical weight under the
coherent part, to which we add the superfluid contribution at $\omega=0$,
is about $1/(1+\lambda)$ of the total weight available
$(\Omega^2_p/8)$
with the remainder, $\lambda/(1+\lambda)$, to be found in the
incoherent part. In the model considered here, which fits the
available data for YBCO$_{6.99}$ and YBCO$_{6.95}$, $\lambda=2.01$
so that only one third of the weight is in the coherent part.
This order of magnitude agrees well with the extensive experimental
results in other cuprates given in Refs.~\onlinecite{tanner,liu}. Note
that coherent and incoherent region are nicely separated over a substantial
frequency range in which the conductivity is small relative to
its value in the quasiparticle peak and in the boson assisted
background. This will lead to a plateau in the integrated optical spectral
weight as a function of the upper limit $\omega$ in the integral
over $\sigma_1(T,\omega)$ which will
in turn lead to an approximate partial or truncated sum rule
on the coherent contribution to the conductivity itself. It is only
this piece which is included in BCS theory and which can be described by
such a theory in cases when it is well separated from the incoherent
background.
 We note that the
addition of impurities, as in the dashed curve in the top
frame of Fig.~\ref{fig:2b}, greatly increases the frequency width
of the quasiparticle peak in $\sigma_1(T,\omega)$ and also fills in the
region between coherent and incoherent part of the conductivity.
While these two contributions are still recognizable as distinct,
they now overlap significantly and cannot as easily be separated.

Finally, but very importantly, in the bottom frame of Fig.~\ref{fig:2b}
we repeat the
curve for $\sigma_1(T,\omega)$ vs $\omega$ for the pure sample
of the top frame of Fig.~\ref{fig:2b} (solid curve) and compare
it with its normal state counterpart (dotted curve). We see that
due to superconductivity, much of the weight
under the Drude peak in the solid curve (superconducting) as
compared with the dotted
curve (normal) has been transferred to the condensate and is not part of
the figure [$\delta$-function at $\omega=0$ in $\sigma_1(\omega,T)$].
 It has also
shifted the incoherent part to higher energies. For an $s$-wave
superconductor the appropriate shift would be twice the gap
as seen in the work of Marsiglio and Carbotte\cite{mars1}
(see their Fig.~11). For the $d$-wave case there is a
distribution of gap values around the Fermi surface and
consequently of upward shifts. This leads to some distortion
of the incoherent part as compared with its normal state value
as can be seen in the dashed curve which is the dotted curve
displaced upwards by $26\,$meV, a value slightly larger than
the gap of $22.3\,$meV and much less than twice the spectral
gap. The difference between dashed and solid curve
is small but not negligible. This shows that in the optical
spectral weight distribution the boson assisted part of the
spectrum is in a first approximation shifted in energy but not
significantly depleted or augmented. The addition of impurities
also have an effect on the incoherent background
as can be seen in the top frame of Fig.~\ref{fig:2b} on
comparison of the solid with the dashed curve. 

\section{Approximate Partial sum rule for the coherent part}
\label{sec:4}

\begin{figure}[t]
  \includegraphics[width=9cm]{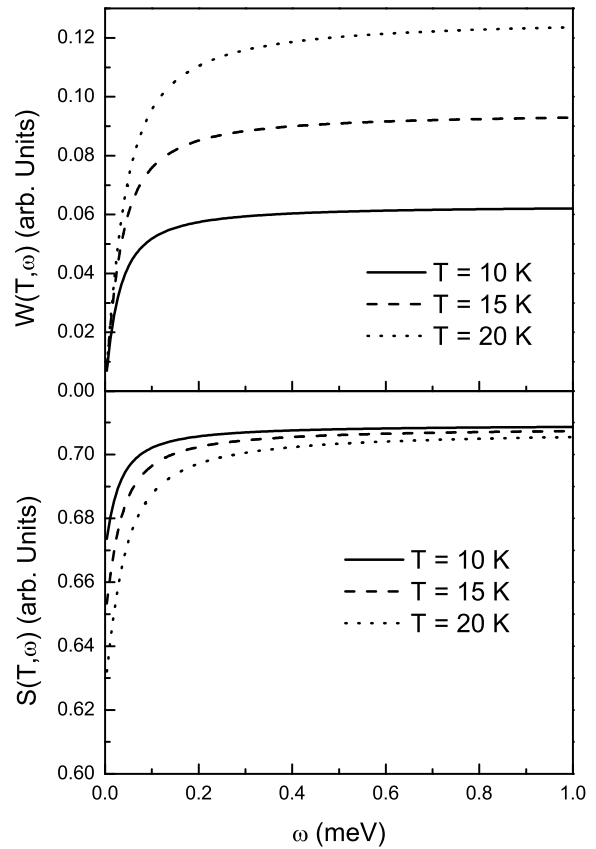}
  \caption{
  The remaining integrated optical spectral weight in the superconducting
  state. Top frame: $W(T,\omega) = \int_{0^+}^\omega d\nu\,
  \sigma_1(T,\nu)$ for values of $\omega$ up to $1\,$meV.
  The temperatures are $10\,$K (solid lines), $15\,$K (dashed lines),
  and $20\,$K (dotted line). The gap is $22.3\,$meV,
  $\Gamma^+=0.003\,$meV and $c=0.2$. Bottom frame:
  $S(T,\omega) = \lim_{\omega\to 0}\omega\sigma_2(T,\omega)+
   2W(T,\omega)/\pi$ in units such that $\lim_{\omega\to\infty}
  S(T,\omega) = 2$.
  }
  \label{fig:3}
\end{figure}

In the top frame of Fig.~\ref{fig:3} we show our theoretical
results for the remaining integrated optical spectral weight
under the real part of
the conductivity $\sigma_1(T,\omega)$ in the superconducting state
up to frequency $\omega$.
By definition $W(T,\omega) = \int_{0^+}^\omega\!d\nu\,\sigma_1(T,\nu)$
where the upper limit of the integral is variable. The data is for
the very pure sample for which coherent and incoherent contributions
are well separated. Results for three temperatures
are shown, namely $T=10\,$K (solid line), $T=15\,$K (dashed line),
and $T=20\,$K (dotted line) and the variable upper limit $\omega$
ranges from zero to $1\,$meV, i.e. only very low frequencies
are sampled, consequently only the coherent quasiparticle contribution
to the conductivity (solid curve in the top frame of Fig.~\ref{fig:s1})
is significantly involved since the incoherent
contribution is almost negligible in this energy range.
Note that already for $\omega\sim 0.4\,$meV a well developed plateau
is seen in each curve, although its magnitude depends on temperature.
$W(T,\omega)$ represents the residual absorption in the microwave
region that remains at low temperatures in the superconducting
state. It decreases with decreasing temperature as more optical
weight is transferred to the condensate.
In our calculations this residual absorption has its origin
in the inelastic scattering associated with thermally
activated bosons which exist at any finite $T$ and which broadens the
quasiparticle contribution. This is in
addition to impurity absorption which is also small, when
$\Gamma^+$ is small. Strictly, at zero temperature
only the impurity absorption remains and this goes to zero
as $\Gamma^+$ goes to zero. We will see later that an
extrapolation to zero temperature of the numerical data for $W(T,\omega)$
gives for the cut off $\omega=1\,$meV, a value of 0.00023 [in units of
$\Omega_p^2/(8\pi)$] which is very small.

In the bottom frame of Fig.~\ref{fig:3} we
show results for a closely related quantity $S(T,\omega)$ vs $\omega$
in units of $\Omega_p^2/(8\pi)$.
In the superconducting state, missing spectral weight under the
real part of the conductivity when compared to its normal state
is found in a delta-function at $\omega=0$ weighted
by the amount in the condensate. In our computer units
the full sum rule which applies when $\sigma_1(T,\omega)$ is integrated to
infinity and the condensate contribution added, is two.
The partial sum up to $\omega$ is
\begin{eqnarray}
  S(T,\omega) &=& \lim_{\omega\to 0}\omega\sigma_2(T,\omega)+
  \frac{2}{\pi}\int\limits_{0^+}^\omega\!d\nu\,\sigma_1(T,\nu)\nonumber\\
  &\equiv& \frac{2}{\pi}\int\limits_0^\omega\!d\nu\,\sigma_1(T,\nu),
  \label{eq:psr}
\end{eqnarray}
and is shown for
the same three temperatures as in the top frame.
Here $\sigma_2(T,\omega)$ is the imaginary part of the
conductivity. When multiplied by $\omega$ its $\omega\to 0$ limit
is proportional to
the inverse square of the London penetration depth which, in
turn, is proportional to the superfluid density.

For an Eliashberg superconductor the expression for the penetration
depth at any temperature $T$ is (in our computer units)
\begin{equation}
  \label{eq:lpdv}
  \frac{1}{\lambda^2_L(T)} = 8\pi\,T
  \sum_{\omega_n}\left\langle\frac{\tilde{\Delta}^2_{{\bf k}'}(\omega_n)}
  {[\tilde{\omega}^2(\omega_n)+\tilde{\Delta}_{{\bf k}'}^2(\omega_n)]^%
{3/2}}\right\rangle'.
\end{equation}
For $T\to 0$ in the constant $\lambda$ model with no impurities we get
\begin{eqnarray}
\frac{1}{\lambda^2_L(T=0)} &=& \frac{8\pi}{1+\lambda}\left\langle
\int\limits_0^\infty\!d\omega\,\frac{\Delta^2\cos(2\phi')}{
[\omega^2+\Delta^2\cos^2(2\phi')]^{3/2}}\right\rangle'\nonumber\\
 & =&
\frac{1}{\lambda^2_{cl}(0)}\left(\frac{1}{1+\lambda}\right).
\label{eq:lpd}
\end{eqnarray}
where we have restored the units
 and $\lambda^{-2}_L(T=0)$ is the usual
value of the London penetration depth.
There are so called strong coupling corrections
to Eq.~\eqref{eq:lpd} (see Ref.~\onlinecite{carb})
but these are small and, in a first approximation, can be neglected.
A physical interpretation of Eq.~\eqref{eq:lpd} is that it is only
the coherent quasiparticle part of the spectral density (Fig.~\ref{fig:spec})
which significantly participates in the condensation.

Returning to the bottom frame of Fig.~\ref{fig:3} we
see that at $\omega\sim 0.4\,$meV a plateau has been reached in
$S(T,\omega)$ vs $\omega$ as well and that, relative to what is the case
for $W(T,\omega)$ in the top frame,
little variation with temperature remains.
Nevertheless, the small amount that is seen will have
consequences as we will describe later.
For now, neglecting this $T$-dependence, the plateau seen
in $S(T,\omega)$ vs $\omega$
implies that an approximate partial sum rule will apply to the
coherent part
of the conductivity by itself, provided the cut off on
$\omega$ is kept small.
This has important implications for the analysis of experiments.
While only approximately $1/(1+\lambda)$ of the optical spectral weight is
involved in this contribution, this piece behaves like a
BCS superconductor. The partial
sum rule which applies, when the cut off $\omega_c$
is kept below the frequency
at which the incoherent part starts to make an important
contribution is
\begin{equation}
  \label{eq:msr}
  S(T,\omega_c) = \lim_{\omega\to 0}\omega\sigma_2(T,\omega)+
  \frac{2}{\pi}\int\limits_{0^+}^{\omega_c}\!d\nu\,\sigma_{1}(T,\nu) \simeq
  \frac{2}{1+\lambda}.
\end{equation}
in the constant $\lambda$ approximation of
Sec.~\ref{sec:2}. In our full Eliashberg calculations
for $T=10\,$K we get $\sim 0.71$ for Eq.~\eqref{eq:msr} instead of
$\sim 2/3$ with $\lambda = 2.01$. It is the existence of the partial
sum rule \eqref{eq:msr} for very pure samples that has allowed
Turner {\it et al.}\cite{turner} to analyze their microwave data
within a BCS formalism without reference to the mid infrared
incoherent contribution. Nevertheless, one has to keep in
mind that this partial sum rule involves only $1/(1+\lambda)$
of the whole spectral weight under the $\sigma_1(T,\omega)$
curve with important consequences on the results derived from such
an analysis.

\begin{figure}[t]
  \includegraphics[width=9cm]{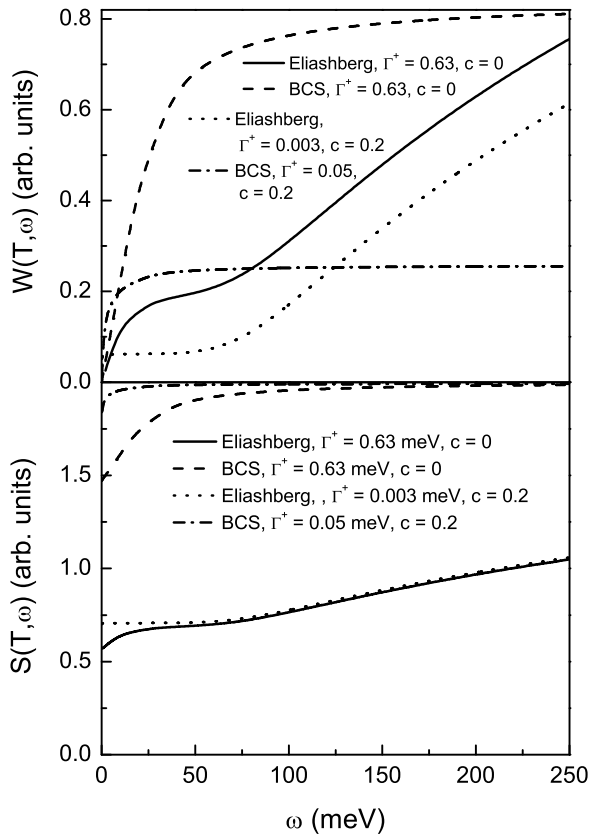}
  \caption{Top frame: The optical spectral weight
    $W(T,\omega) = \int_{0^+}^\omega d\nu\,\sigma_1(T,\nu)$
    as a function of the upper limit $\omega$.
    Two curves apply to BCS and
    two correspond to Eliashberg calculations. In one case the
    unitary limit $(c=0)$ is used with
    $\Gamma^+ = 0.63\,$meV (solid curve
    for Eliashberg, dashed for BCS). The dotted curve is similar
    but for $\Gamma^+ = 0.003\,$meV and $c=0.2$ in Eliashberg
    theory and the dash-dotted is for $\Gamma^+ = 0.05\,$meV,
    $c=0.2$ in BCS.
    Bottom frame: the same as the top frame but now the
    sum $S(T,\omega) = \lim_{\omega\to 0}\omega\sigma_2(T,\omega)+%
    (2/\pi)\int_{0^+}^\omega d\nu\,\sigma_1(T,\nu)$ is shown.
    In both frames the temperature $T=10\,$K and the
    $d$-wave gap amplitude is the same for Eliashberg and
    BCS calculations.
    }
  \label{fig:4}
\end{figure}

For the pure case considered here the cutoff $\omega_c$
in \eqref{eq:msr} is well defined.
This is further illustrated in Fig.~\ref{fig:4} where we
show once more $W(T,\omega)$ (top frame) and $S(T,\omega)$ (bottom frame) but
now for an extended frequency range up to $250\,$meV for the case
$T=10\,$K only. We also show, for comparison, additional BCS results
and results for a second set of impurity parameters. 
The solid and dotted curves in both
frames are $W(T,\omega)$ and $S(T,\omega)$ for an Eliashberg superconductor
with $\Gamma^+=0.63\,$meV, $c=0$ and $\Gamma^+=0.003\,$meV, $c=0.2$,
respectively. The dashed and dash-dotted curves are for a
BCS superconductor with $\Gamma^+=0.63\,$meV, $c=0$ and
$\Gamma^+=0.05\,$meV, $c=0.2$. We first note that for the purer
Eliashberg case (dotted curve)
the plateau in both, $W(T,\omega)$ and $S(T,\omega)$
identified in Fig.~\ref{fig:3} extends to $\omega\simeq 50\,$meV.
Clearly, any value of frequency between $\omega\simeq 0.4\,$meV
and $50\,$meV will do for $\omega_c$ in Eq.~\eqref{eq:msr} and
a partial sum rule is well defined but for the less pure case (solid
curve) a plateau is not as well defined.
In both cases, however, the
increase beyond the plateau value of $\sim 0.7$ towards saturation
is rather slow and even at $\omega=250\,$meV $S(T,\omega)$ is
still well below 2. This feature reflects directly the large
energy scale involved in the boson exchange mechanism we have
used. This behavior is in sharp contrast to BCS. For the
dash-dotted curve $S(T,\omega)$ is already close to two at
$\omega\simeq 25\,$meV while for the less pure case (dashed curve)
the rise to two is slower and distributed over a larger energy
scale of the order $\sim 100\,$meV. In as much as impurities
strongly affect such scale estimates they are not fundamental
to the superconductivity itself. If, in our Eliashberg
calculations, we look only at the initial rise to its
plateau value $(\sim 0.7)$, the scales involved are
different again, $\sim 1\,$meV and $\sim 50\,$meV respectively.

\section{Relation between residual absorption and penetration depth}
\label{sec:5}

\begin{figure}[t]
  \includegraphics[width=9cm]{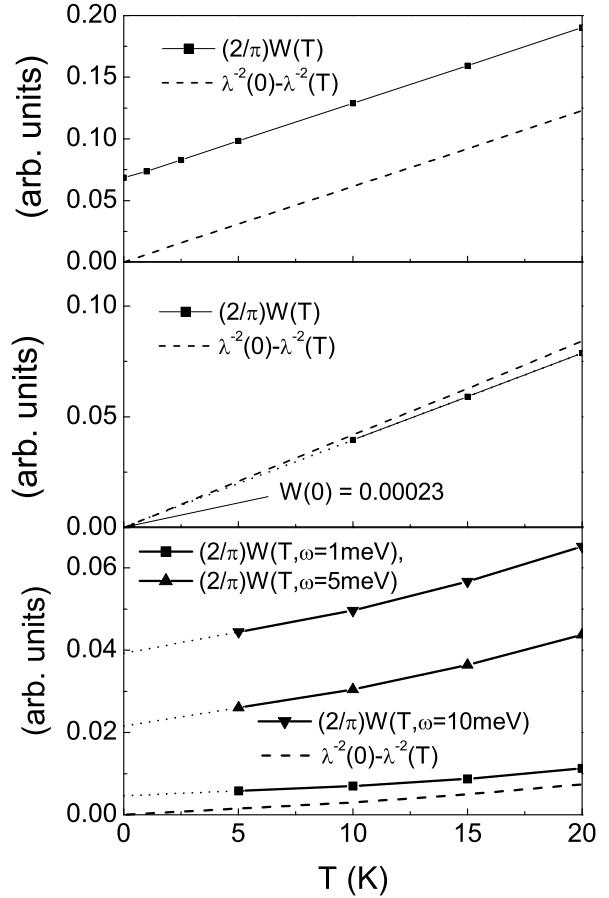}
  \caption{Top frame: comparison of $(2/\pi)W(T)$ vs $T$ (dotted
    curve) with $\lambda^{-2}(0)-\lambda^{-2}(T)$ (dashed curve)
    for a BCS $d$-wave superconductor with the gap amplitude set
    at $\Delta=24\sqrt{2}\,$meV and with
    $\Gamma^+ = 0.1\,$meV and $c=0.3$. The
    lines are parallel to each other. The superfluid density
    goes to zero at $T=0$ while the remaining area under
    the real part of the conductivity
    goes to a finite value (residual absorption).
    Middle frame: same comparison as in the top frame
    for an
    Eliashberg superconductor modeled for
    YBa$_2$Cu$_3$O$_{6.99}$, with $\Gamma^+ = 0.003\,$%
meV and $c=0.2$. The curve for $(2/\pi)W(T)$ extrapolates
    to a very small value as $T\to 0$ and
    the two curves are not quite parallel. Bottom frame:
    same as for the middle frame but with $\Gamma^+=0.63\,$meV and
    $c=0.2$. Three different cut offs in $W(T,\omega)$ are used.
    }
  \label{fig:5}
\end{figure}

We next turn to the relationship between the temperature dependence
of the residual absorption and the penetration depth. This is
illustrated in Fig.~\ref{fig:5} which has three frames. The top
frame presents BCS results and is for comparison with the two
other frames which are based on Eliashberg solutions. The central
frame has impurity parameters $\Gamma^+=0.003\,$meV and $c=0.2$.
The bottom frame is for a less pure sample with $\Gamma^+=0.63\,$meV
and $c=0.2$ 
and illustrates how impurities change the results.
In the top frame, the dashed curve is the difference in
superfluid density $\lambda^{-2}(0)-\lambda^{-2}(T)$ as a
function of temperature $T$ up to $20\,$K for a BCS superconductor
with gap $\Delta = 24\sqrt{2}\,$meV, $\Gamma^+=0.1\,$meV, and
$c=0.3$. These parameters were chosen only for the purpose of
illustration. Turner {\it et al.}\cite{turner} considered the
optical spectral weight concentrated in the microwave region
of an ortho-II YBCO$_{6.5}$ sample and the temperature dependence of
$W(T)$ that is obtained from consideration of the microwave
region only. They found it to extrapolate to a finite
value at $T=0$ (zero temperature residual absorption) while at
the same time $W(T)$ parallels the temperature dependence found
for the penetration depth. In our solid curve (top frame of
Fig.~\ref{fig:5} we have integrated
$\sigma_1(T,\omega)$ to get $W(T,\omega)$ up to $1\,$meV and find a
curve for $W(T)$ which is parallel to the dashed curve for the
penetration depth but indeed does not extrapolate to zero at
$T=0$. Note that in a BCS model for pure samples the ordinary
FGT sum rule applies even if only the microwave region is considered
and so the solid and dashed curves are parallel.
This is no longer the case in Eliashberg theory as shown
in the center frame of Fig.~\ref{fig:5}. There the dashed and solid
curves are not quite parallel with the dashed curve showing a
slightly steeper slope. Also, the solid curve
extrapolates to a finite though very small value at $T=0$. This is expected
since the impurity content in this run is very small. This
case corresponds closely to the YBCO$_{6.99}$ sample considered
in Fig.~4 of Turner {\it et al.}\cite{turner}
The slight difference in slope between solid
and dashed curve can be understood in terms of our result for
$S(T,\omega_c)$ given in the bottom frame of Fig.~\ref{fig:3}. We
have already noted that at $\omega=1\,$meV, the cutoff used in
evaluation of $W(T,\omega_c)$ (solid curve, center frame of Fig.~\ref{fig:5})
there remains a small temperature dependence
to the saturated value of $S(T,\omega_c)$.
This means that $S(T,\omega_c)$ in this region 
is slightly smaller at $T=20\,$K (dotted curve in the bottom frame
of Fig.~\ref{fig:3}) than it is at $T=10\,$K (solid curve). This
slight deviation from the partial sum rule
embodied in our Eq.~\eqref{eq:msr}
leads immediately to the difference in slope seen in the center
frame of Fig.~\ref{fig:5} between $W(T,\omega_c)$ and the penetration depth.

In the bottom frame of Fig.~\ref{fig:5} we show results for
$\Gamma^+=0.63\,$meV and $c=0.2$. In this case the coherent
and incoherent contribution to $\sigma_1(T,\omega)$ (see Fig.~\ref{fig:2b},
top frame, dotted curve although this curve is for $c=0$) are not
as well separated as in the pure case
and $W(T,\omega)$ vs $\omega$ does not show as
clear a plateau which would allow the formulation of a partial sum
rule on the coherent part alone.
Nevertheless, we do note that for
$\omega_c = 1\,$meV, $2W(T,\omega_c)/\pi$ (solid
squares) is nearly parallel to the dashed curve for the
penetration depth. If, however, $\omega_c$ is increased to
$5\,$meV (solid up-triangles), or $10\,$meV (solid down-triangles)
this no longer holds. This result can be traced to the fact
that no real temperature and cut off independent plateau
is reached in these cases. Thus, there is
no partial sum rule which can be
applied on $W(T,\omega)$ and an analysis as performed by
Turner {\it et al.}\cite{turner} on very high purity samples appears
not to be possible. This case may correspond better to the
relatively dirtier film data.\cite{corson}
Note in particular, the residual absorption at zero temperature
depends now strongly on the cutoff frequency chosen for the partial
sum rule. In our example (bottom frame of Fig.~\ref{fig:5}) the
residual absorption increases almost linearly with increasing
cutoff frequency.

We turn next to the zero temperature value of the residual absorption
and its
impurity dependence. Eq.~\eqref{eq:lpdv} applies but now we wish
to consider impurities so that $\tilde{\omega}_n$ is not simply
$\tilde{\omega}_n = \omega_n(1+\lambda)$ in the constant $\lambda$
model. Instead, we must use
\begin{equation}
  \label{eq:iom}
  \tilde{\omega}(\omega+i0^+) = \omega(1+\lambda)+
  i\pi\Gamma^+\frac{\Omega(\omega)}{c^2+\Omega^2(\omega)},
\end{equation}
which needs to
be solved self consistently for $\tilde{\omega}(\omega+i0^+)$.
For $\omega=0$, we can write $\tilde{\omega}(\omega+i0^+)=i\gamma$
with
\begin{equation}
  \label{eq:gam}
  \gamma = \pi\Gamma^+\frac{\Omega(i\gamma)}{c^2+\Omega^2(i\gamma)}
\end{equation}
and $\Omega(i\gamma)$ is given by Eq.~\eqref{eq:dos}. Evaluating
$\Omega(i\gamma)$ gives
\begin{equation}
  \label{eq:8}
  \gamma = \pi\Gamma^+\frac{\frac{2\gamma}{\pi\Delta(1+\lambda)}
    \ln\left(\frac{4\Delta(1+\lambda)}{\gamma}\right)}{c^2+
    \left(\frac{2\gamma}{\pi\Delta(1+\lambda)}\right)^2
    \ln^2\left(\frac{4\Delta(1+\lambda)}{\gamma}\right)}.
\end{equation}
This transcendental equation for $\gamma$, the zero frequency scattering
rate at zero frequency, is to be solved numerically for
any value of $c$. Results can be found in Refs.~\onlinecite{schach2}
and \onlinecite{schach1} for the case $\lambda=0$.
What is found is that $\gamma/c$ increases with $\Gamma^+$
and, for a given value of $\Gamma^+$ decreases
rapidly with $c$. At $c=0$ we get the approximate, but very useful
relation
\begin{equation}
  \label{eq:gamap}
  \gamma = 0.63\sqrt{\pi\Gamma^+\Delta(1+\lambda)}.
\end{equation}
Note, this is the
same expression as in Hirschfeld and Goldenfeld\cite{hirschf2} except
that it contains an additional factor of $(1+\lambda)$. In terms of
$\gamma$ we can get an approximate expression for the zero
temperature London penetration depth including impurities. Returning
to Eq.~\eqref{eq:lpdv} we need to replace $\tilde{\omega}_n$
by $\omega_n(1+\lambda)+\gamma$ to get\cite{schach1}
\begin{widetext}
\begin{eqnarray}
  \label{eq:lpdc}
  \frac{1}{\lambda^2_L(0)} &=& 8\pi\frac{1}{1+\lambda}
  \int\limits_0^{2\pi}\!d\phi\int\limits_0^\infty\!d\omega\,
  \frac{\Delta^2\cos^2(2\phi)}{\left[\left(\omega+\frac{\gamma}{1+\lambda}
  \right)^2+\Delta^2\cos^2(2\phi)\right]^{3/2}}\\
  \label{eq:lpdd}
  &\simeq&\frac{1}{\lambda^2_{cl}(0)}\left\{1-
   \frac{2}{\pi}K\left[\frac{i}{\gamma}\Delta(1+\lambda)\right]\right\},
\end{eqnarray}
\end{widetext}
where $K(x)$ is the elliptic integral of the first kind.
The approximation made to get the last equality,
Eq.~\eqref{eq:lpdd}, is not very accurate but has the
the important advantage that it is analytic and simple. It
gives
\begin{equation}
  \label{eq:6}
  \lambda^{-2}_L(0) \simeq \lambda^{-2}_{cl}(0)\left[
    1-\frac{2\gamma(1+\lambda)}{\pi\Delta}\ln\left(\frac{4\Delta
    (1+\lambda)}
      {\gamma}\right)\right].
\end{equation}
In a BCS model $(\lambda=0)$ this gives
in the limits $T\to 0$ and $\omega\to\infty$
\begin{eqnarray}
  W(T=0,\omega\to\infty) \equiv W(0)
  &=& \int\limits_{0^+}^\infty\!d\omega\,\sigma_1(0,\omega)\nonumber\\
    &=& \frac{\gamma}{\Delta}\ln
      \left(\frac{4\Delta}{\gamma}
      \right).
  \label{eq:5}
\end{eqnarray}
Exact numerical results for $W(0)$ based on Eq.~\eqref{eq:lpdc}
\begin{figure}[t]
  \includegraphics[width=9cm]{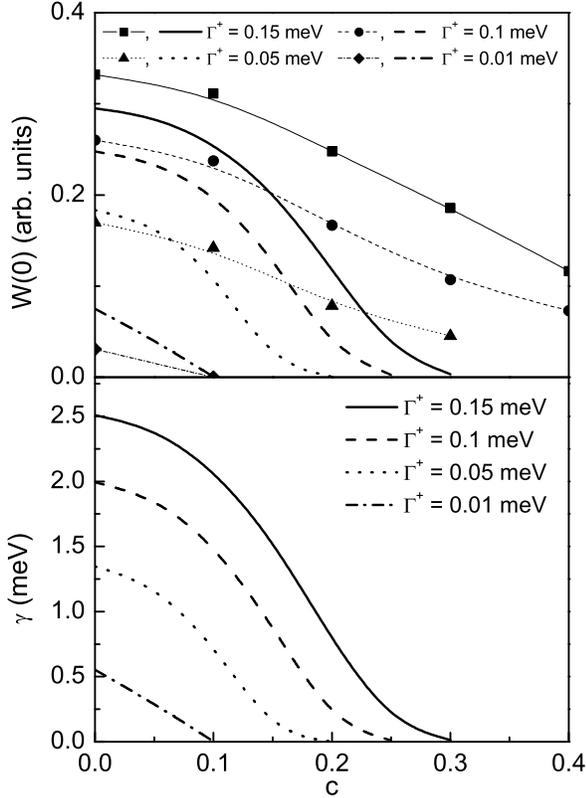}
  \caption{Top frame: the $T\to 0$ limit of the remaining optical
    spectral weight $W(0)=\int_{0^+}^\infty d\omega\,\sigma_1(T=0,\omega)$
    as a function of the impurity potential strength $c$ for
    various values of $\Gamma^+$.
    The heavy continuous curves are the approximation
    $W(0)\simeq (\gamma/\Delta)\ln(4\Delta/\gamma)$ while
    the light curves with solid squares ($\Gamma^+=0.15\,$meV),
    solid circles ($\Gamma^+=0.1\,$meV), solid triangles
    ($\Gamma^+=0.05\,$meV), and solid diamonds ($\Gamma^+=0.01\,$meV)
    are exact results. The bottom frame gives the zero
    frequency value of the effective scattering in the superconducting
    state, $\gamma(c)$ as a function of $c$.
    }
  \label{fig:6}
\end{figure}
with $\lambda=0$ are compared with those based on Eq.~\eqref{eq:5}
in the top frame of Fig.~\ref{fig:6}.
We see that Eq.~\eqref{eq:5} is
qualitatively but not quantitatively correct. In the bottom frame we
show the corresponding values of $\gamma(c)$ vs $c$ for the convenience
of the reader. It is clear that the
residual absorption due to the coherent part of the charge carrier
spectral density does depend significantly on impurity content.
In a real superconductor we have additional absorption at
$T=0$ coming from the incoherent, boson assisted background
which enters when $\omega$ in the upper limit of the
defining integral for $W(T,\omega)$ is made to span energies
in the infrared region of the spectrum.

\section{Missing area}
\label{sec:6}

\begin{figure}[t]
  \includegraphics[width=9cm]{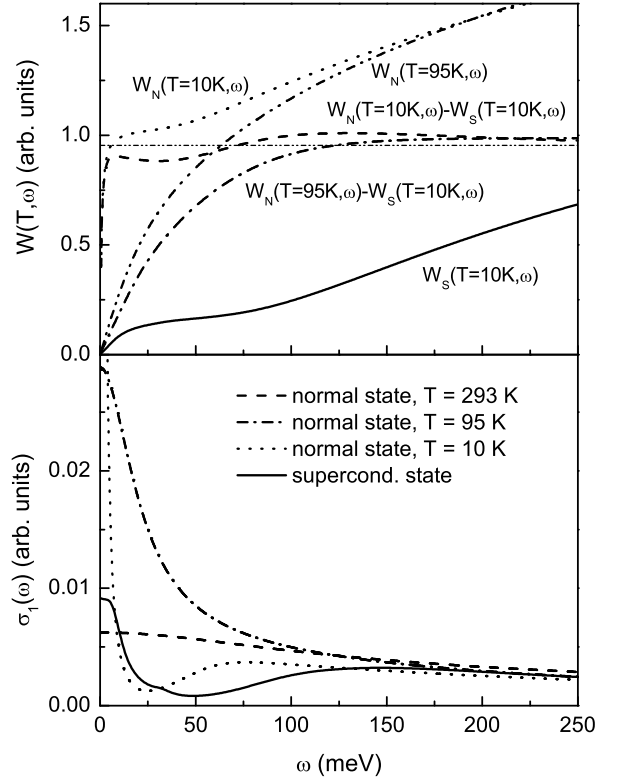}
  \caption{Top frame: optical spectral weight
$W(\omega,T) = \int_{0^+}^\omega\!d\nu\,\sigma_1(\nu)$ for various cases
as a function of $\omega$. The dotted (dash-double dotted) curve is
for the normal state at $T=10\,$K ($T=95\,$K), the solid curve for
the superconducting state at $T=10\,$K. The dashed (dash-dotted)
curve is the difference curves between superconducting and normal state
($\tilde{\Delta}(\omega) =0$ in the Eliashberg equations)
at $T=10\,$K ($T=95\,$K).
The approach of the difference in area
to its saturated large $\omega$ value
depends significantly on the temperature used for
the subtracted normal state. The thin dash-double dotted horizontal
line is the value of the penetration depth.
Bottom frame: it shows the real part
of the conductivity for the normal state at $T=293\,$K (dashed curve),
$T=95\,$K (dash-dotted curve), $T=10\,$K (dotted curve), and for the
superconducting state a $T=10\,$K (solid curve). All curves are
for YBCO$_{6.95}$ with the impurity parameters set to
$\Gamma^+=0.63\,$meV and $c=0$.}
\label{fig:7}
\end{figure}

The FGT sum rule implies
that the missing optical spectral weight under the real part of the
conductivity in the superconducting state appears as a
delta function contribution at the origin proportional
to the superfluid density. It depends on temperature and on
impurity content. Increasing $T$ and/or $\Gamma^+$ decreases the
superfluid density. In the top frame of Fig.~\ref{fig:7} we show
our results for the remaining integrated
optical spectral weight $W(T,\omega)$ as a function
of $\omega$ up to $250\,$meV for a sample with $\Gamma^+=0.63\,$meV
and $c=0$. We have done similar calculations for a clean sample but
there is no qualitative difference. The solid curve is for the
superconducting state at $T=10\,$K and is to be compared with
the dotted curve which is for the normal state at the same temperature.
We see a great deal of missing spectral weight between
these two curves with $W_N(T,\omega)$ rising much faster
at small $\omega$ than $W_S(T,\omega)$ and it
is rising to a much higher value. The difference
$W_N(\omega,T=10\,{\rm K})-W_S(\omega,T=10\,{\rm K})$
(dashed curve) is the amount
of optical spectral weight between $(0^+,\omega)$ that has been
transferred to the superfluid condensate.
As we see, the dashed
curve rapidly grows within a few meV to a
value close (but not quite)
to the asymptotic value it assumes at $\omega = 250\,$meV.
After this the remaining variation is small
but there is a shallow minimum around $30\,$meV with a corresponding
broad and slight peak around $100\,$meV which is followed
by a small gradual decrease still seen at $250\,$meV. These features
can be understood in detail when the frequency dependence of
$\sigma_1(T,\omega)$ is considered. The relevant curves to be
compared are the dotted (normal) and solid (superconducting)
in the bottom frame of Fig.~\ref{fig:7}. Both are at $10\,$K.
The curves cross at 3 places on the frequency axis. Above the
first crossing at $\omega_1\approx 8\,$meV the difference in
the integrated area decreases till $\omega_2\approx 32\,$meV
at which it begins to increase. Finally, at the third
crossing $\omega_3\approx 130\,$meV it begins to decrease
again towards its value at $250\,$meV. These features are the
direct result of the shift in incoherent background towards
higher energies due to the opening up of the
superconducting gap. The area between the dotted and solid
lines that falls
between $\omega_2$ and $\omega_3$ is made up slowly at higher
frequencies. This feature would not be part of BCS theory in
which case the energy scale for the optical weight which significantly
participates in the condensate is set as a few times the gap
$\Delta$\cite{dolgov} and the saturated value is reached from
below rather than from above. In our theory the existence of the
incoherent background effectively increases this scale to
much higher energies, the scale set by the bosons involved,
 although the amount of spectral weight
involved is very small.\cite{homes1,sant} We note that at
$\omega=250\,$meV the missing area curves $W_N(T=10\,{\rm K},%
\omega)-W_S(T=10\,{\rm K},\omega)$ and $W_N(T=95\,{\rm K},%
\omega)-W_S(T=10\,{\rm K},\omega)$ of the top frame of
Fig.~\ref{fig:7} are still about 2.5\% higher than the value
indicated for the penetration depth (thin dash-double dotted line)
which is obtained directly from the imaginary part of the
optical conductivity.

In an actual experiment it is not possible to access
the normal state at low temperatures so that  $W_N(\omega,T=10\,{\rm K})$
cannot be used to compute the difference with $W_S(\omega,T=10\,{\rm K})$.
Usually $W_N(\omega,T=95\,{\rm K})$ is
used instead. This is shown as the dashed-double dotted curve in
the top frame of Fig.~\ref{fig:7} which is seen to merge with the
dotted curve only at large values of $\omega$. Because in our
theoretical work, the inelastic scattering at $T=T_c$ is large with
a scattering rate of the order $2T_c$ or so, the corresponding
optical spectral weight in $\sigma_1(T,\omega)$ is shifted to higher
energies. Consequently, $W_N(\omega,T=T_c)$ rises much more slowly
out of $\omega=0$ than does $W_N(\omega,T=10\,{\rm K})$ and the
difference curve $W_N(\omega,T=95\,{\rm K})-W_S(\omega,T=10\,{\rm K})$
(dash-dotted curve) reflects this. It merges with the
dashed curve only for $\omega\stackrel{>}{\sim} 200\,$meV. Thus,
making use of $W_N(\omega,T=T_c)$ rather than $W_N(\omega,T=10\,{\rm K})$
makes a considerable difference in the estimate of the $\omega$
dependence of the missing area. None of the structure seen in the
dashed curve remains in the dash-dotted curve and much information
on separate coherent and incoherent contributions is lost,
although the curve still approaches its $\omega\to\infty$ limiting
value from above. From this point of view, it is
the dashed curve which is fundamental but it is not directly
available in experiments. If an even higher temperature had been
used for the normal state, say around room temperature, the
frequency at which the difference $W_N(\omega)-W_S(\omega)$
would agree with the penetration depth is pushed to very high energies
well beyond the $250\,$meV range shown in the top frame of Fig.~%
\ref{fig:7}. The reason for this is clear when the bottom frame
of this same figure is considered. What is shown is the real part of the
conductivity for four cases: the normal state at $T=293\,$K
(dashed curve), at $T=95\,$K (dash-dotted curve), and at
$T=10\,$K (dotted curve). Increasing the normal state temperature
shifts a lot of spectral weight to higher energies and can even
make the difference $W_N-W_S$ negative for small $\omega$.

We stress again that individual $W(T,\omega)$ curves show no
saturation as a function of $\omega$ in the range shown. This
is characteristic of the high $T_c$ oxides and resides in the
fact that $I^2\chi(\omega)$, the electron-boson exchange
spectral density,
extends to very high energies. This is fundamental to an
understanding of the optical properties in these materials
and is very different from the electron-phonon case. In that
instance there is a maximum phonon energy $\omega_D$ never
larger than about $100\,$meV and hence the curve for $W(T,\omega)$
would reach saturation at a much smaller energy than in our work.
This observation provides strong evidence against
solely a phonon mechanism for superconductivity in the
oxides.

\begin{figure}[t]
  \includegraphics[width=9cm]{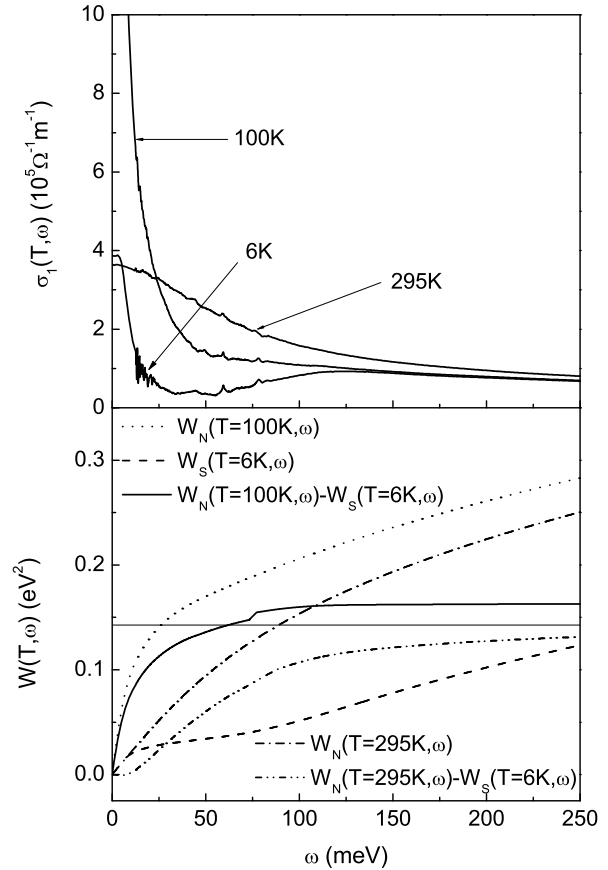}
  \caption{Top frame: Experimental data for the real part of the
optical conductivity, $\sigma_1(\omega)$ vs $\omega$ and
various temperatures for an
optimally doped Bi2212 single crystal as it was reported by
Tu {\it et al.}\protect{\cite{tu}} The data has been augmented
by theoretical data\protect{\cite{schach8}} in the energy range
$0<\omega\le 12.\,$meV. Bottom frame: Optical spectral weight
$W(\omega,T) = \int^{\omega}_{0^+}\!d\nu\,\sigma_1(\nu,T)$ vs $\omega$
as calculated from the experimental data shown in the top frame
of this figure. The dashed line is for $T=6\,$K (superconducting
state), the dotted line for $T=100\,$K, and the dash-dotted line
for $T=295\,$K. Presented are also the differences
$W_N(\omega,T=295\,$K)-$W_S(\omega,T=6\,$K) (dash-double dotted
line) and $W_N(\omega,T=100\,$K)-$W_S(\omega,T=6\,$K) (solid line).
The thin, solid horizontal line represents the theoretical
value $(\pi/2)\lim_{\omega\to 0}\omega\sigma_2(\omega,T=6\,$K).
    }
  \label{fig:9}
\end{figure}

To aid this discussion we added Fig.~\ref{fig:9} which, in its
top frame, shows the experimental data for the real part
of the optical conductivity, $\sigma_1(T,\omega)$
reported by Tu {\it et al.}\cite{tu} in an optimally doped
Bi$_2$Sr$_2$Ca Cu$_2$O$_{8+\delta}$ (Bi2212) single crystal
for three temperatures, namely,
$T = 6$, $100$, $295\,$K. The experimental data has been augmented
by theoretical data\cite{schach8} in the frequency region
$0<\omega\le 12.4\,$meV derived from best fits to experiment. This
graph is to be compared with the bottom frame of Fig.~\ref{fig:7}.
The bottom frame of Fig.~\ref{fig:9} presents the
corresponding optical spectral
weight $W(\omega,T)$ calculated from the experimental
$\sigma_1(\omega,T)$ data. The results follow closely similar
theoretical curves presented in the top frame of Fig.~\ref{fig:7}.
In particular, $W_S(\omega,T=6\,$K) does not develop a 
well defined plateau
around $50\,$meV as we found it for optimally doped YBCO$_{6.95}$
single crystals [solid line in the top frame of Fig.~\ref{fig:7},
labeled $W_S(T=10\,$K$,\omega)$]. Finally, the differences
$W_N(\omega,T=100\,{\rm K})-W_S(\omega,T=6\,{\rm K})$ (solid
line) and  $W_N(\omega,T=295\,{\rm K})-W_S(\omega,T=6\,{\rm K})$
(dash-double dotted line) are shown in this graph. We also
included the theoretical value for
$(\pi/2)\lim_{\omega\to 0}\omega\sigma_2(\omega,T=6\,{\rm K})$
as a thin, solid horizontal line found from a fit to experimental
data. The first difference is still far away from this limit but
approaches it from above, as expected from our previous discussion,
while the second approaches this limit from below. This  analysis
of experimental data supports our theoretical results in a rather
impressive way.

\section{Conclusions}
\label{sec:7}

In a pure BCS superconductor at zero temperature with no impurities
the entire optical spectral weight under the real part of the
conductivity will vanish as it is all transferred to the superfluid
density which contributes a $\delta$-function at $\omega=0$ to the
real part of $\sigma(\omega)$. When impurities are present the
superfluid density at $T=0$ is reduced from its clean limit value
and some spectral weight remains under $\sigma_1(\omega)$
which implies some absorption even at zero temperature. The
situation is quite different for a superconductor which
shows a pronounced incoherent
background scattering which can be
modeled reasonably well in Eliashberg theory be
it $s$- or $d$-wave. In both cases it is mainly the coherent part of the
electron spectral density which contributes to the condensate. The electron
spectral function still has a $\delta$-function part broadened
by the interactions at any finite energy away from the Fermi
energy but the amount of weight under this part is $1/(1+\lambda)$,
where $\lambda$ is the mass enhancement parameter for the
electron-boson exchange interaction. The remaining spectral
weight $\lambda/(1+\lambda)$ is to be found in incoherent, boson
assisted tails. Another way of putting this is that at
zero temperature in a pure system the
superfluid density is related to the renormalized plasma
frequency with $m^\ast$ replacing the bare electron mass
$(m^\ast/m = 1+\lambda)$ in contrast to the total plasma frequency
which involves the bare mass $m$. The
incoherent, boson assisted tails in $\sigma_1(T,\omega)$
do not contribute much to the condensate and in
fact remain pretty well unaffected in shape and optical weight
by the transition to the
superconducting state but they are shifted upwards due to the opening
up of the superconducting gap. This shift implies that when
one considers the missing optical spectral weight under the
conductivity which enters the condensate, the energy scale
for this readjustment is not set by the gap scale but rather
by the scale of the maximum exchanged boson energy. Also
it is expected that the value of the penetration depth which
corresponds to the saturated value of the missing area is
approached from above when the conductivity is integrated
to high energies.

\section*{Acknowledgment}
 
Research supported by the Natural Sciences and Engineering
Research Council of Canada (NSERC) and by the Canadian
Institute for Advanced Research (CIAR). J.P.C. thanks D.M.~Broun
for discussions. The authors are grateful to Drs. C.C.~Homes
and J.J.~Tu for providing their original experimental data
for analysis.


\begin{thebibliography}{99}
  \bibitem{mars1}F.~Marsiglio and J.P.~Carbotte, Aust.\ J.\ Phys.
    {\bf 50}, 975 (1997), and {\bf 50}, 1011 (1997).
  \bibitem{ferrell}R.A.~Ferrell and R.E.~Glover, Phys.\ Rev.
    {\bf 109}, 1398 (1958).
  \bibitem{tinkham}M.~Tinkham and R.B.~Ferrell, \prl {\bf 2},
    331 (1959).
  \bibitem{vM}H.J.A.~Molegraaf, C.~Presura, D.~van der Marel,
    P.H.~Kes, and M.~Li, Science {\bf 295}, 2239 (2002).
  \bibitem{schach8}E.~Schachinger and J.P.~Carbotte, J.~%
    Phys.~Studies (L'viv, Ukraine) {\bf 7}, 209 (2003);
    E.~Schachinger and J.P.~Carbotte, in: {\it Models and Methods of
       High-TC Superconductivity: some Frontal Aspects, Vol. II},
       ed.: J.K.~Srivastava and S.M.~Rao,
       Nova Science, Hauppauge NY (2003),  pp. 73.
  \bibitem{quinla}S.M.~Quinlan, P.J.~Hirschfeld, and
    D.J.~Scalapino, \prb {\bf 53}, 8575 (1996).
  \bibitem{hirschf}P.J.~Hirschfeld, S.M.~Quinlan, and
    D.J.~Scalapino, \prb {\bf 55}, 12\ 742 (1997).
  \bibitem{quinlb}S.M.~Quinlan, D.J.~Scalapino,and
    N.~Bulut, \prb {\bf 49}, 1470 (1994).
  \bibitem{hirschf1}P.J.~Hirschfeld, W.O.~Putikka, and D.J.~Scalapino,
    \prb {\bf 50}, 10250 (1994).
  \bibitem{BCS6}F.\ Marsiglio and J.P.\ Carbotte, in Handbook on
    Superconductivity: Conventional and Unconventional, eds.
    K.H.\ Bennemann and J.B.\ Ketterson (Springer, Berlin, 2003),
    pp.~233-345.
  \bibitem{carb}J.P.~Carbotte, Rev.\ Mod.\ Phys. {\bf 62}, 1027 (1990).
   \bibitem{tanner}D.B.~Tanner, H.L.~Liu {\it et al.}, Physica B
    {\bf 244}, 1 (1998).
  \bibitem{liu}H.L.~Liu, N.A.~Quijada, {\it et al.}, J.\ Phys.:
    Condens.\ Matter {\bf 11}, 239 (1999).
 \bibitem{turner}P.J.~Turner, R.~Harris, S.~Kamal, M.E.~Hayden,
    D.M.~Broun, D.C.~Morgan, A.~Hosseini, P.~Dosanjh, G.~Mullins,
    J.S.~Preston, R.~Liang, D.A.~Bonn, and W.N.~Hardy,
    \prl {\bf 90}, 237005 (2003).
  \bibitem{corson}J.~Corson, J.~Orenstein, Seongshik Oh, J.~%
  O'Donnell, and J.N.~Eckstein, \prl {\bf 85}, 2569 (2000).
  \bibitem{resabs}E.~Schachinger and J.P.~Carbotte, \prb {\bf 67},
    134509 (2003).
  \bibitem{no7}G.\ Grimvall, The Electron-Phonon Interaction
     in Metals, (North-Holland, New York, 1981).
  \bibitem{pines1}A.\ Millis, H.\ Monien, and D.\ Pines,
     \prb {\bf 42}, 167 (1990).
  \bibitem{pines2}P.\ Monthoux and D.\ Pines, \prb {\bf 47},
     6069 (1993); \prb {\bf 49}, 4261 (1994); \prb {\bf 50},
     16015 (1994).
  \bibitem{schach4}J.P.\ Carbotte, E.\ Schachinger, and D.N.\ Basov,
    Nature (London) {\bf 401}, 354 (1999).
  \bibitem{schach5}E.\ Schachinger and J.P.\ Carbotte, \prb
    {\bf 62}, 9054 (2000).
  \bibitem{schach7}E.\ Schachinger, J.P.\ Carbotte, and D.N.\ Basov,
    Europhys.\ Lett. {\bf 54}, 380 (2001).
  \bibitem{schach6}E.~Schachinger and J.P.~Carbotte, Physica C {\bf 364},
    13 (2001).
  \bibitem{varma1}C.M.\ Varma, Int.\ J.\ Mod.\ Phys. {\bf 3},
     2083 (1989).
  \bibitem{varma2}P.B.\ Littlewood, C.M.\ Varma, S.\ Schmitt-Rink,
     and E.\ Abrahams, \prb {\bf 39}, 12371 (1989).
  \bibitem{varma3}C.M.~Varma, P.B.~Littlewood, S.~Schmitt-Rink,
     E.~Abrahams, and A.E.~Ruckenstein, \prl {\bf 63}, 1996 (1989);
     {\it ibid.} {\bf 64}, 497 (1990).
  \bibitem{mcmillan}W.L.~McMillan and J.M.~Rowell, \prl {\bf 19},
    108 (1965).
  \bibitem{mcmillan1}W.L.\ McMillan and J.M.\ Rowell, in
    Superconductivity, ed. R.D. Parks (Marcel Dekker Inc., New York,
    1969), p. 561.
 \bibitem{mars4}F.\ Marsiglio, T.\ Startseva, and J.P.\ Carbotte,
    Phys.\ Lett.\ A {\bf 245}, 172 (1998).
  \bibitem{bourges}Ph.\ Bourges, Y.\ Sidis, H.F.\ Fong, B.\ Keimer,
     L.P.\ Regnault, J.\ Bossy, A.S.\ Ivanov, D.L.\ Lilius, and
     I.A.\ Aksay, in High Temperature Superconductivity, eds. S.E.\
     Barnes, {\it et al.}, CP483 (American Institute of Physics,
     Amsterdam, 1999), p. 207.
  \bibitem{puchkov}A.V.\ Puchkov, D.N.\ Basov, and T.\ Timusk,
    J.\ Phys.: Condens.\ Matter {\bf 8}, 10049 (1996).
  \bibitem{homes}
     C.C.\ Homes, D.A.\ Bonn, R.\ Liang, W.N.\ Hardy, D.N.\ Basov,
     T.\ Timusk, and B.P.\ Clayman, \prb
     {\bf 60}, 9782 (1999).
  \bibitem{tu}J.J.\ Tu, C.C.\ Homes, G.D.\ Gu, D.N.\ Basov,
    and M.\ Strongin, \prb {\bf 66}, 144514 (2002).
  \bibitem{sorella}S.~Sorella, G.B.~Martins, F.~Becca, C.~Gazza,
     L.~Capriotti, A.~Parola, and E.~Dagotto, \prl {\bf 88},
     117002 (2002).
  \bibitem{hoss}A.~Hosseini, R.~Harris, S.~Kamal, P.~Dosanjh,
    J.~Preston, R.~Liang, W.N.~Hardy, and D.A.~Bonn, \prb {\bf 60},
    1349 (1999).
  \bibitem{schach2}E.~Schachinger and J.P.~Carbotte, \prb
  {\bf 64}, 094501 (2001).
  \bibitem{lee}P.A.~Lee, Phys.~Lett. {\bf 71}, 1887 (1993).
  \bibitem{schach1}E.~Schachinger and J.P.~Carbotte, \prb {\bf 65},
    064514 (2002).
  \bibitem{schur}I.~Sch\"urrer, E.~Schachinger, and J.P.~Carbotte,
    Physica C {\bf 303}, 287 (1998); J.\ Low Temp.\ Phys. {\bf 115},
    251 (1999).
  \bibitem{hirschf2}P.J.~Hirschfeld and N.~Goldenfeld, \prb {\bf 48},
    4219 (1993).
  \bibitem{dolgov}A.E.~Karakozov, E.G.~Maksimov, and
  O.V.~Dolgov, Solid State Commun. {\bf 124}, 119 (2002).
  \bibitem{homes1}C.C.~Homes, S.V.~Dordevis, D.A.~Bonn,
  R.~Liang, and W.N.~Hardy, cond-mat/0303506 (unpublished).
  \bibitem{sant}A.F.~Santander-Syro, R.P.M.S.~Lobo,
  N.~Bontemps, Z.~Konstantinovic, Z.Z.~Li, and
  H.~Raffy, Europhys. Lett. {\bf 62}, 568 (2003).
\end{thebibliography}
\end{document}